# Galvatron: Automatic Distributed Training for Large Transformer Models

**Esmail Gumaan**
Faculty of Computer Science and Information Technology
University of Sana'a
Yemen, Sana'a
esmailG231601.ue.ye.edu

## Abstract

Training multi-billion to trillion-parameter language models efficiently on GPU clusters requires leveraging multiple parallelism strategies. We present Galvatron, a novel open-source framework (dubbed "Optimus-Megatron" in the implementation) that dynamically combines data parallelism, tensor model parallelism, and pipeline parallelism to optimize training throughput. Built atop PyTorch and integrating NVIDIA's Megatron-LM and Microsoft's DeepSpeed, Galvatron automatically selects and adjusts parallelism strategies in real-time based on model architecture, hardware, and training dynamics. This paper details Galvatron's key features—automatic hybrid parallelism selection, layer-wise and phase-wise strategy optimization, and runtime adaptation—and contrasts them with existing static frameworks. We describe the system's technical stack, including its use of DeepSpeed's ZeRO and NCCL communication, and an in-depth implementation overview of its core modules (profilers, strategy selector, parallelism manager). We then illustrate how Galvatron can be seamlessly integrated into existing training pipelines with minimal code modifications, providing companies a plug-and-play solution for efficient large-model training. Finally, we situate Galvatron in context with related efforts (NVIDIA Megatron-LM, Microsoft DeepSpeed, Google GShard, Meta FairScale, etc.), highlighting how it advances the state-of-the-art in distributed deep learning. We provide references to the GitHub repository and relevant literature throughout.

**GitHub Code:** https://github.com/Esmail-ibraheem/Galvatron

**Galvatron** is an open-source system that automates distributed training for Transformer-based deep learning models. It addresses the challenges of efficiently training large models across multiple GPUs by automatically determining optimal hybrid parallelism strategies. This paper presents an in-depth look at Galvatron's concept, architecture, and implementation. We describe Galvatron's core components – a hardware/model profiler, a parallelism search engine, and a distributed runtime – and explain how they work together to maximize training throughput. We also discuss the codebase structure and technical details, providing guidance for organizations to integrate Galvatron into their own model training pipelines. The goal is to offer a clear understanding of Galvatron's design and usage, enabling companies and researchers to adopt its techniques for efficient large-scale model training.

## 1   Introduction

Training large language models (LLMs) such as GPT-3 and BERT at scale has become increasingly challenging due to the immense computational and memory requirements. To fit models with billions or trillions of parameters across multiple GPUs, practitioners combine parallelization strategies **(data parallelism, model parallelism, pipeline parallelism)**. Existing frameworks like NVIDIA's Megatron-LM and Microsoft's DeepSpeed provide support for distributed training, but they rely on static, manually-configured parallelism plans that users must set before training. This static approach requires expert tuning and may not be optimal for all training phases or hardware conditions, leading to suboptimal efficiency.

**Galvatron** addresses these limitations by introducing a dynamic hybrid parallelism framework called Optimus-Megatron in the codebase. Rather than fixing a parallel strategy upfront, Galvatron can automatically identify and deploy the optimal combination of data, model, and pipeline parallelism at each stage of training. It continuously monitors training metrics (throughput, memory usage, etc.) and adjusts the parallelism strategy in real time as the model progresses, without user intervention. By automating parallelism selection and adaptation, Galvatron aims to maximize hardware utilization and training speed while simplifying the user experience.

**Contributions** of Galvatron include:

- Automatic Parallelism Selection: Determining an optimal hybrid parallel strategy (mix of data, tensor, pipeline parallelism) for the given model and hardware without manual tuning.

- Dynamic Strategy Adjustment: Adapting the parallelism configuration on the fly based on real-time metrics (e.g. GPU utilization, communication overhead, convergence rate), allowing different strategies at different training phases.

- Layer-Wise and Phase-Wise Optimization: Using layer-specific parallelism (e.g. tensor parallel for attention layers, data parallel for feed-forward layers) and adjusting strategy between early and late training (throughput-oriented vs stability-oriented).

- Seamless Integration and Usability: Building on existing high-performance libraries (Megatron-LM, DeepSpeed) so that users can adopt Galvatron with minimal code changes, abstracting away the low-level details behind a simple interface.

By tackling the inefficiencies of static parallelism schemes, Galvatron **paves the way for faster and more scalable training** of next-generation LLMs on diverse hardware setups. In the following sections, we discuss Galvatron's design and implementation in detail, outline the technical stack it builds upon, and explain how organizations can leverage this framework in practice. We also compare Galvatron's approach to related systems and prior research in **distributed deep learning** and **hybrid parallelism**.

## 2 Key Features and Innovations

**Galvatron** introduces a form of **Dynamic Hybrid Parallelism** that extends traditional "hybrid parallelism" with runtime adaptability. In conventional hybrid parallel training, one might combine data parallelism, model (tensor) parallelism, and pipeline parallelism in a fixed configuration for a given run. For example, a user might decide to train a 24-layer Transformer by splitting layers into 4 pipeline stages, using 2-way tensor model parallelism within each stage, and replicating this setup across 2 data-parallel replicas, and these choices remain constant throughout training. Galvatron, by contrast, makes these decisions **dynamic and context-aware**:

- **Dynamic Strategy Selection:** Galvatron employs an **"AI-driven" strategy selector** that can choose between parallelism modes and their degrees on the fly. Initially, it uses profiled information about the model and hardware to set up a good starting plan (e.g. how many pipeline stages, how many GPUs per model-parallel group). Then during training, it continually monitors metrics and can **switch strategies** if it detects bottlenecks or shifts in resource usage. For instance, Galvatron might **start with model parallelism** for large attention layers in early epochs and then **transition to data parallelism** in later epochs when gradient communication dominates costs. This adaptive switching is done automatically by the framework, unlike static frameworks where such changes would require manual reconfiguration and restarting training.

- **Layer-Specific Parallelism Optimization:** Rather than applying one parallelism scheme uniformly, Galvatron can apply **different parallel strategies to different parts of the model**. The underlying insight is that layers have different computation and memory profiles. In a Transformer, **self-attention layers** are compute-heavy and memory-intensive – these benefit from **tensor model parallelism** (splitting the layer's matrix multiplications across GPUs). In contrast, **feed-forward layers** (MLPs) are lighter per parameter and might benefit more from data parallelism (to increase the effective batch size and statistical efficiency). Galvatron's strategy selector and model profiler identify such opportunities, enabling a **mix of parallelism within one model**: e.g. attention layers distributed in model-parallel fashion while MLP layers replicated data-parallel across GPUs. This layer-wise optimization is determined dynamically based on profiling of layer shapes and compute costs.

- **Training Phase Adaptation:** Galvatron recognizes that the optimal strategy may change over the course of training. Early in training, throughput and exploration might be prioritized – for example, using **pipeline parallelism with many micro-batches** to maximize GPU utilization. Later in training, when models approach convergence, stability and consistent updates might matter more – favoring simpler data parallelism or fewer pipeline stages to reduce stale updates. The framework can **dynamically reconfigure** aspects like pipeline depth or micro-batch sizing as these needs evolve. An example scenario described in the documentation: starting with 4 pipeline stages and then **reducing to 3 stages mid-training** upon detecting imbalance in stage workloads, and shifting emphasis to data-parallel in final epochs for convergence.

- **Automated Tuning with AI Guidance:** Unlike prior systems that rely on fixed heuristics or exhaustive search, Galvatron's design integrates AI-driven decision making to guide parallelism selection. In principle, this could involve using a learned model (e.g. a reinforcement learning agent or neural cost model) that predicts the best strategy given the current state (model size, hardware stats, training progress). The repository suggests that the framework can incorporate RL or predictive models to continuously improve strategy choices. While the current implementation primarily uses a rules-based decision tree and dynamic programming search, the system's architecture is built to accommodate more sophisticated AI optimization in the future, making it **future-proof for new hardware and model trends**.

- **Improved Usability and Abstraction:** A major distinguishing feature of Galvatron is the emphasis on **user-**friendliness despite the complexity under the hood. In static hybrid parallel setups, users often must manually configure many low-level parameters: how to partition the model, number of pipeline stages, assignment of layers to devices, data-parallel replication factors, etc.. Galvatron hides this complexity. **The user only needs to provide the model and dataset**, and optionally high-level constraints like a target batch size or enable/disable dynamic adaptation flag. The framework's Parallelism-Manager then autonomously orchestrates the rest. From a user's perspective, training with Galvatron can be as simple as calling a provided train(model, data) utility or integrating the **Parallelism-Manager** into their training loop, without having to modify the model code for parallel execution. **This plug-and-play** design lowers the barrier for organizations to leverage advanced parallelism techniques, since no deep expertise in distributed systems is required.

## 3   Technical Stack and System Architecture

Galvatron is implemented as a **Python-based framework** building on top of the PyTorch deep learning library. It leverages several state-of-the-art systems in distributed training as its foundation, integrating them under its dynamic parallelism layer:

**PyTorch & NCCL:** At the lowest level, Galvatron uses PyTorch's distributed capabilities and NVIDIA's NCCL for communication. This provides efficient tensor operations on GPUs and fast collectives (all-reduce, all-gather, etc.) for gradient and parameter synchronization. The choice of PyTorch means Galvatron can interface easily with popular model implementations (like HuggingFace Transformers or Megatron models) and use CUDA kernels optimized for Transformer operations.

**NVIDIA Megatron-LM:** The project incorporates components of Megatron-LM [2], NVIDIA's framework for training multi-billion parameter Transformers. Megatron-LM provides **intra-layer model parallelism** (also known as tensor parallelism) where weight matrices are partitioned across GPUs to split the compute of large layers. It also uses highly optimized **ring-allreduce** algorithms for syncing gradients across data-parallel replicas. By building on Megatron-LM's core, Galvatron reuses these low-level optimizations and data structures (such as model parallel group communicators and sharded weight initialization) instead of reinventing them. **The megatron** module in the Galvatron repository contains modifications and extensions to Megatron-LM's code to support dynamic behavior (for example, hooks that allow changing parallel groups during training).

**Microsoft DeepSpeed:** Galvatron also integrates with **DeepSpeed**, particularly for its **ZeRO** optimizations [3] and pipeline parallel engine. DeepSpeed's ZeRO (Zero Redundancy Optimizer) partitions optimizer states and gradients across GPUs, substantially reducing memory usage in data-parallel training. This enables **scaling to very large models (trillion+ parameters)** by overcoming GPU memory limits. Galvatron can make use of DeepSpeed by initializing models with `deepspeed.initialize()` if enabled, thereby combining Galvatron's dynamic strategy selection with DeepSpeed's memory-efficient training. Additionally, DeepSpeed provides a flexible pipeline parallelism implementation that Galvatron can control. By interfacing with DeepSpeed, Galvatron ensures compatibility with a widely-used industry solution and inherits features like gradient accumulation, distributed checkpointing, and more, out-of-the-box.

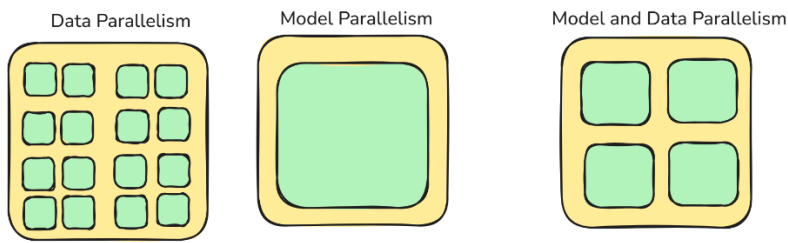

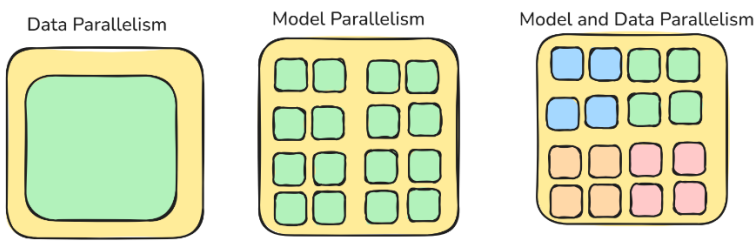

Figure 1: Model Parallelism V.s Data Parallelism

**Hardware/Cluster Awareness:** The framework assumes a cluster of GPUs connected with high-speed links (in a single node or across nodes). It automatically initializes the PyTorch distributed process group (with NCCL backend for GPUs) if not already done. It detects the number of GPUs, available memory per GPU, and the interconnect type (e.g. PCIe vs NVLink) via its Hardware Profiler. This information is used to guide initial parallel strategy (e.g., on NVLink-connected multi-GPU node, model parallelism might be cheaper communication-wise than across nodes on slower interconnects).

**Operating Environment:** Galvatron's repository is configured as a standalone Python package. The code is organized to be importable (it includes an __init__.py and uses the megatron namespace for core components). It likely requires a CUDA-enabled PyTorch installation and has been tested on Linux with GPU clusters. The project is open-sourced under the MIT License, making it easy for companies to adopt and modify. In terms of dependencies, aside from PyTorch and DeepSpeed, it uses standard Python libraries for threading and logging (as seen by use of threading.Lock and Python logging in the code). It also likely depends on NVIDIA's communication libraries (which come with PyTorch's NCCL support) and possibly the HuggingFace or Megatron models for the actual model definitions (the repository includes example scripts for GPT and BERT pretraining which presumably use Megatron's model definitions).

**System Architecture:** Architecturally, Galvatron adds a "dynamic optimization layer" on top of the static parallel training engines of Megatron-LM/DeepSpeed. As shown in the Figure 2.

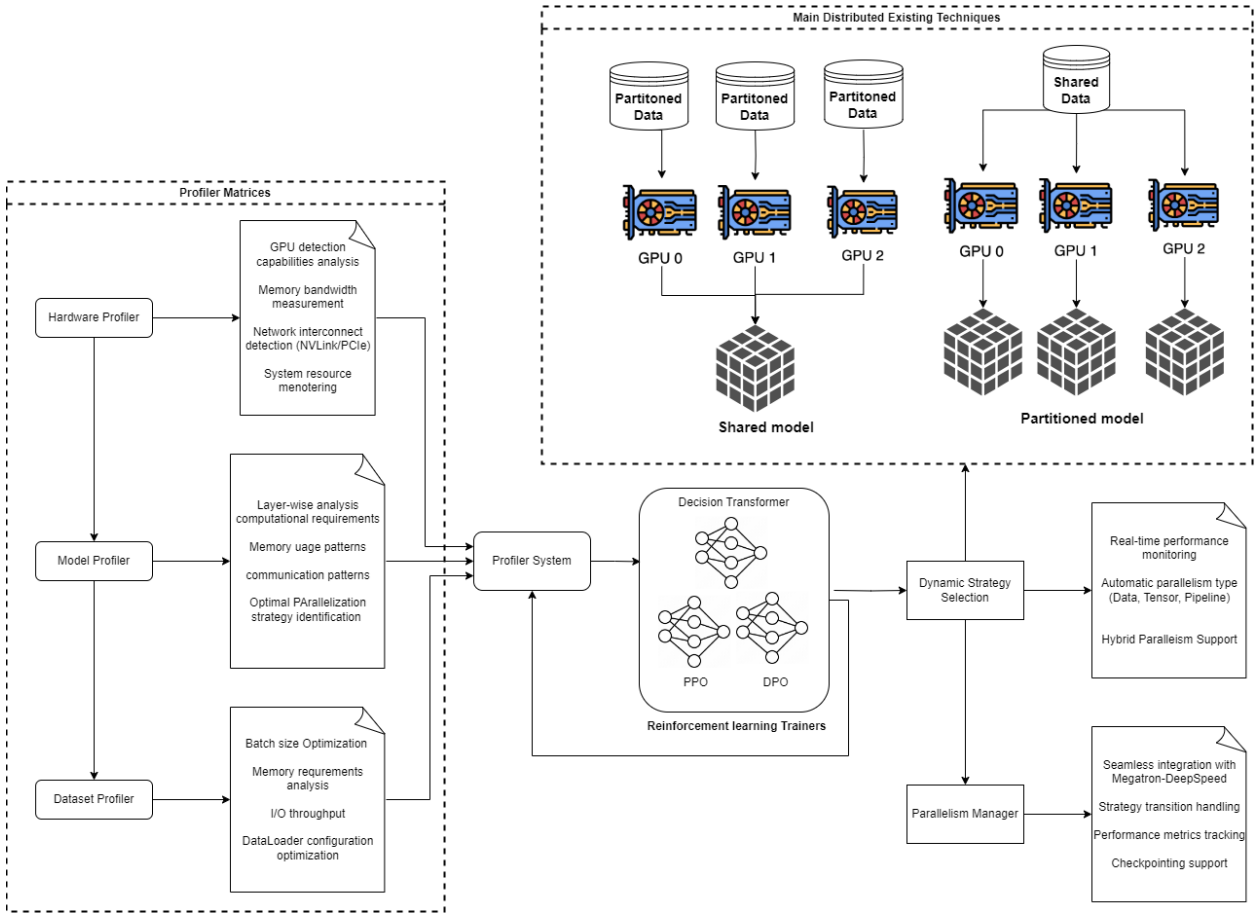

Figure 2: Galvatron System Architecture

conceptually shows Galvatron sitting between the user's training script and the underlying distributed training operations. The key components of Galvatron's architecture include:

**Hardware Profiler:** This module inspects the hardware environment at startup. It queries the number of GPUs, their memory capacity, and network topology (e.g., checks if GPUs are on the same node with NVLink or across nodes over Ethernet/InfiniBand). The profiler may use PyTorch's CUDA APIs (like `torch.cuda.get_device_properties`) to get memory and device info, and NCCL or system queries for interconnect bandwidth. This info helps decide, for example, if pipeline parallelism (which is sensitive to inter-GPU bandwidth) is feasible or if one should limit to intra-node tensor parallel groups.

**Model Profiler:** This component analyzes the neural network model structure. It can compute the total parameter count and possibly the per-layer shapes and compute cost (e.g., FLOPs or memory per forward pass for each layer). By identifying layers or modules that are particularly large or computationally heavy, the model profiler highlights where model parallelism would be beneficial. For instance, if certain layers have an extraordinarily high parameter count (like the embedding layer or a huge feed-forward layer), the profiler tags those for potential partitioning. It also helps determine how many pipeline stages might be appropriate by looking at layer count and balancing work per stage(e.g., evenly distribute transformer layers across stages).

**Dataset Profiler:** This module evaluates the dataset and training configuration. Key factors include the size of the dataset, the global batch size, and possibly the memory footprint of one sample or batch. A large available dataset or the ability to use very large batch sizes might push toward more data parallelism (since data parallel efficiency improves with larger batch sizes). Conversely, if batch size is constrained by memory, other forms of parallelism (model/pipeline) must compensate to reach high device utilization. The dataset profiler might also benchmark data loader throughput to ensure the parallel strategy (especially pipeline parallelism which increases effective batch latency) aligns with input pipeline capabilities.

**Dynamic Strategy Selector:** This is the central "brain" of Galvatron. It **synthesizes the information from the hardware, model, and dataset profilers to generate an optimized parallelism plan**. The plan is essentially a configuration: how many pipeline stages, how many way tensor parallelism, how many data-parallel replicas (and by extension micro-batch size if using pipelining). In the **Discovery Phase** (initialization), the strategy selector uses heuristic rules and possibly a search algorithm to choose an initial hybrid strategy before training starts. Notably, the Galvatron paper describes using a decision tree to prune the search space and then a dynamic programming algorithm to find an optimal strategy given a performance model. During the **Optimization Phase**, the strategy selector continues to operate, but now it takes in real-time training metrics gathered during the **Monitoring Phase**. These metrics can include: **throughput (samples/sec), GPU utilization, memory usage, communication overhead time, load imbalance between GPUs, and convergence rate (loss improvement per step)**. Galvatron's training loop calls the strategy selector (via the `ParallelismManager.step()` API) periodically with updated metrics. If the metrics indicate suboptimal performance (for example, one stage of pipeline is running much slower than others causing bubbles, or GPUs have unused memory that could allow bigger batches), the strategy selector decides if a change in strategy is warranted. It can then output a new parallelism configuration (for instance, switch from 8-way data parallel to 4-way data + 2-way tensor parallel if communication overhead is too high, etc.). Implementing such a switch involves coordination with the Parallelism Manager to repartition the model or adjust pipeline divisions on the fly. The strategy selector in code (see `DynamicStrategySelector` class) encapsulates this logic and maintains the current strategy state.

**Parallelism Manager:** The ParallelismManager is the **runtime orchestrator** that applies the strategies selected by the above component. It interfaces directly with the training process and underlying frameworks. On initialization, the ParallelismManager uses the chosen plan to **set up the distributed training environment**: it defines process groups for model parallelism, splits the model's layers across pipeline stages (if pipeline parallelism is used) by assigning subsets of layers to different devices, and initializes any necessary **infrastructure for switching** strategies later (such as keeping a copy of the model or checkpointing mechanisms that allow repartitioning). During training, the ParallelismManager receives updates from the strategy selector. When a decision is made to change the parallelism configuration, the manager handles the transition – for example, pausing training briefly, **migrating or resharding** the model (if going from tensor-parallel to data-parallel, it would gather shards of weights; if adding pipeline stages, it would redistribute layers across GPUs), and then resuming training with the new configuration. This is a complex process that involves synchronizing all workers, possibly saving and loading model states. Galvatron's implementation strives to make transitions efficient; for instance, it may perform these adjustments at certain **intervals or safe points** (like after completing an epoch or when the model is checkpointed) to minimize disruption. The ParallelismManager also encapsulates utilities for micro-batch handling in pipeline parallelism (splitting a batch into micro-batches for the pipeline and recombining outputs), and it ensures that the optimizer and learning rate scheduler are aware of the parallel training context (so that things like gradient averaging or update frequency are correctly accounted for). In the code, ParallelismManager holds references to the model, optimizer, scheduler, and the strategy selector and communication optimizer, tying everything together. It maintains a lock and state flags to manage concurrency issues during strategy transition.

**Communication Optimizer:** In addition to managing what parallel strategy to use, Galvatron includes a **CommunicationOptimizer** module to improve efficiency within a given strategy. Modern distributed training can overlap communication (like gradient all-reduce) with computation, and fuse smaller communication messages into larger ones to reduce overhead. The CommunicationOptimizer likely monitors communication time vs compute time (the communication_overhead metric passed into `manager.step()`) and toggles optimizations such as tensor fusion, communication scheduling, or gradient accumulation lengths to hide latency. In the code snippet, we see the CommunicationOptimizer being initialized with options `enable_fusion=True`, `enable_overlap=True`, indicating it will try to fuse communication operations and overlap them with computation. This component complements the parallelism strategy: for example, if using data parallel with many GPUs (which entails large all-reduce overhead), the comm optimizer will ensure NCCL calls are fused and overlapped to minimize impact. If pipeline parallelism is used, it might adjust the number of micro-batches (since pipeline with more micro-batches naturally overlaps communication of gradients with forward/backward passes). By actively managing communication, Galvatron can further close the performance gap between an ideal strategy and its real-world execution.

All these components work in concert during a training run. The **workflow** can be summarized in three phases (as outlined in the documentation):

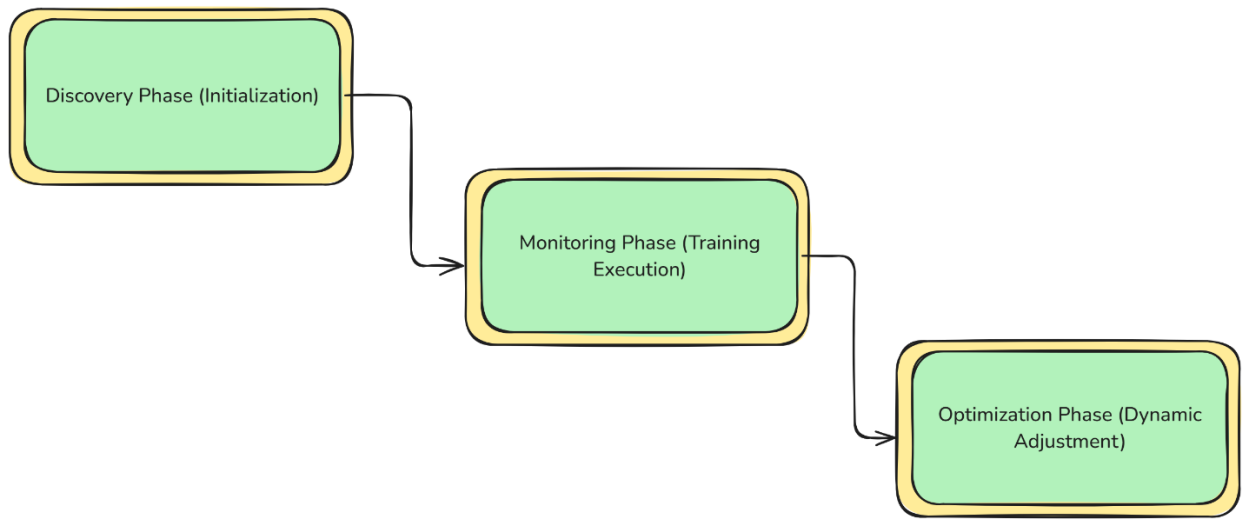

Figure 3: Galvatron System Workflow (high-overview)

1. **Discovery Phase (Initialization)**: The hardware, model, and data are profiled, and the Dynamic Strategy Selector computes an initial parallelism plan. The Parallelism Manager then initializes the distributed environment accordingly (setting up process groups, partitioning the model). This could be seen as finding a good starting point in the huge search space of possible parallelism configurations.

2. **Monitoring Phase (Training Execution):** Training proceeds using the chosen strategy. During this phase, Galvatron collects runtime statistics. The training loop is augmented with calls to the ParallelismManager (and thereby the strategy selector) every so many iterations (or every iteration, depending on config) with metrics like throughput per GPU, recent gradient variance or loss improvement (convergence rate), communication time, etc.. These metrics are aggregated by the profilers or the manager itself. The overhead of monitoring is kept low, using lightweight instrumentation or PyTorch's built-in profilers.

3. **Optimization Phase (Dynamic Adjustment):** At certain intervals, Galvatron evaluates if a change in strategy could improve performance. The Dynamic Strategy Selector uses the collected metrics to detect bottlenecks or inefficiencies. For example, if it observes that GPUs are under-utilized (perhaps waiting on communication), it might decide to reduce the degree of data parallelism (so that each GPU has more work per batch) or increase tensor parallelism (to use more GPUs per layer and thus shorten training time per batch). If it sees memory headroom on GPUs, it might increase the micro-batch size to better utilize memory bandwidth. Or if some pipeline stage is significantly slower, it could redistribute layers (e.g., move a heavy layer to a different stage or split a stage). When a new plan is chosen, the Parallelism Manager carries out the transition. After adjustment, the training continues with the new settings. This loop of monitor -> analyze -> adjust continues until training completes. If no issues are found, Galvatron may stick with the initial plan, so it only intervenes when profitable.

By the end of training, Galvatron will also handle **cleanup** – ensuring any altered process groups are properly destroyed and that final model weights (potentially sharded across devices) are consolidated for saving or evaluation. This design allows long-running training jobs to self-tune their parallel configuration, akin to how a database optimizes queries during execution or an OS scheduler balances processes – an analogy that highlights Galvatron's system-oriented approach to deep learning.

## 4 Implementation Details

The Galvatron GitHub repository provides a concrete implementation of the above architecture in roughly 2.6K lines of Python code. The code is structured into modules corresponding to the described components, along with integration hooks into Megatron/DeepSpeed training pipelines:

- **Code Organization:** The repository contains a core/ directory with key modules (hardware_profiler.py, model_profiler.py, dataset_profiler.py, strategy_selector.py, parallelism_manager.py, etc.) and a megatron/ directory which likely includes modified Megatron-LM code (such as definitions of model parallel groups, distributed data loader, and possibly a modified training loop to call Galvatron's manager). Example scripts like pretrain_gpt.py and pretrain_bert.py show how to initialize training with Galvatron's dynamic parallelism enabled. The design is such that existing Megatron-DeepSpeed training scripts can be adapted by importing Galvatron's components.

- **ParallelismManager and Training Loop**: One key integration point is the model setup and training loop. In Megatron-LM (and DeepSpeed), there is usually a function to initialize the model, optimizer, and distributed training groups. Galvatron injects itself here: as shown in the documentation's code snippet, after constructing the model, it creates instances of **ModelProfiler**,

**HardwareProfiler**, and **ParallelismManager**, then attaches a **DynamicStrategySelector** to the model. This effectively augments the model with the ability to decide strategies. If DeepSpeed is enabled, Galvatron allows DeepSpeed to initialize as usual (which sets up ZeRO, etc.), but importantly the parallel groups (like model parallel groups for Megatron) are already initialized by Galvatron's manager prior to DeepSpeed init. This ensures DeepSpeed is aware of the parallel decomposition Galvatron chose (for example, it won't try to split the model again differently). During training, the ParallelismManager provides a method step(metrics_dict) that the training loop calls every iteration (or periodically) to update the dynamic strategy selector with new metrics . The metrics can be collected by timing the iterations (to get throughput), examining the loss (to compute recent convergence rate), measuring communication time (e.g., using DeepSpeed's timers or torch.profiler), etc. The snippet in the README shows the typical usage in a training loop, where after each loss = train_step(batch), the code does manager.step({...metrics...}). Internally, ParallelismManager.step() likely increments a counter and when sufficient iterations have passed (based on a monitoring interval threshold), it calls the strategy selector's logic to possibly trigger an adjustment. If a decision to change is made, the manager coordinates the change (which might involve re-partitioning the model and optimizer states). The training loop is oblivious to these internal changes except perhaps a slight pause when they occur. After training completes, the loop calls manager.cleanup() to finalize the process.

- **Profilers:** The Hardware, Model, and Dataset profilers are implemented as simple classes that gather stats. For example, HardwareProfiler might run at initialization and store a dictionary of hardware info (GPU count, memory per GPU, links). ModelProfiler might iterate over model.modules() to count parameters and possibly identify layer types/sizes. These are likely used by the DynamicStrategySelector to inform its decisions. The dataset profiler could analyze the dataset or dataloader (if available) to suggest an initial global batch size or number of data parallel replicas appropriate for the data size Since these profiling tasks are relatively fast, they add minimal overhead at startup.

- **DynamicStrategySelector Implementation:** The strategy selector's code is where the decision algorithm resides. In the repository, strategy_selector.py defines DynamicStrategySelector which contains methods to evaluate performance metrics and update the strategy. It also defines data structures for representing a parallelism configuration, likely via an enum or class ParallelismConfig containing fields like strategy_type (data, tensor, pipeline or combinations) and specific parameters (e.g., pipeline_parallel_size, tensor_parallel_degree, micro_batch_size). We saw an example: if total parameters > 1e9, the initial config chosen was to use pipeline parallelism across all GPUs with a micro-batch of 32. Such rules come from the idea that very large models need model/pipeline parallelism to fit in memory. The strategy selector likely includes logic for other cases (if model is smaller, maybe pure data parallel is fine). During training, the selector might use thresholds or learned models to decide if, say, communication overhead is above X%, then try reducing data parallel degree. Or if GPU utilization is below some percentage, try increasing pipeline micro-batches or shifting to model parallel for more compute per GPU. The exact algorithm could be complex; the original research suggests a search in a huge space of configurations. The code may implement a simplified version or a heuristic approach for real-time use. Nonetheless, a core aspect is that it maintains the current config (current_config) and can output a new ParallelismConfig when needed.

- **Handling Transitions:** One of the toughest implementation challenges is moving from one parallelism strategy to another mid-training. Galvatron handles this likely by leveraging model checkpoints. The ParallelismManager might save a checkpoint of the model/optimizer state in a consistent way, then reorganize the model distribution and load the state back. For example, going from model-parallel to data-parallel involves concatenating weight shards and broadcasting them; going from data-parallel to model-parallel means slicing weights and sending shards to different GPUs. The manager's code contains functions _split_model or _combine_model (not shown here, but implied) to do these transformations. Pipeline parallel transitions might involve changing how layers are grouped, which could be done by reinitializing the pipeline groups and informing each process which layers to hold. The threading lock in ParallelismManager suggests they carefully ensure that only one strategy change happens at a time (important if multiple triggers coincide). In practice, the system might restrict major changes to occur between epochs or after a certain minimum interval to avoid oscillations. The CommunicationOptimizer also likely plays a role during transitions, perhaps flushing any in-flight communications and adjusting to new world sizes.

- **Logging and Debugging:** Given the complexity, the code likely has logging (as indicated by self.logger = logging.getLogger('ParallelismManager')) to record decisions and actions. This would help in analyzing how the strategy changed over time and in debugging any performance issues.

## 5 Literature Review and Related Work

Techniques for large-scale distributed training of neural networks have been evolving rapidly. Galvatron's approach builds upon and extends ideas from several **prior systems and research projects** in model parallelism, data parallelism as shown in Figure [1], and hybrid strategies:

**5.1 NVIDIA Megatron-LM:** Shoeybi et al. (2019) introduced Megatron-LM [2], a framework for training multi-billion parameter language models by combining intra-layer model parallelism with data parallelism. In Megatron-LM, each transformer layer's large matrix multiplications are split across multiple GPUs (model parallel), allowing much larger models to fit in memory. Meanwhile, multiple such model-parallel groups are run in data parallel to scale out to many GPUs. Megatron-LM achieved impressive scaling for GPT-like models (e.g., the original Megatron paper trained a 8.3 billion parameter model on 512 GPUs) and introduced optimized communication (ring all-reduce) to keep data parallel synchronization efficient. However, Megatron-LM by itself requires the user to specify the parallelism dimensions (e.g., 8-way model parallel, 64-way data parallel, etc.) at launch and these remain fixed. Galvatron leverages Megatron-LM's efficient model parallel primitives but goes further by automating the choice and adjustment of parallelism over time. In essence, Galvatron can be seen as an evolution of Megatron-LM that adds a dynamic control loop on top of Megatron's static scheme, enabling better adaptability to different model sizes and training stages.

**5.2 Microsoft DeepSpeed:** DeepSpeed is a deep learning optimization library focusing on enabling "extreme-scale" model training. A key contribution of DeepSpeed is the ZeRO (Zero Redundancy Optimizer) [2, 3] family of techniques (Rajbhandari et al. 2020) which partition model states across data-parallel processes to drastically reduce memory duplication. With ZeRO, DeepSpeed was able to train models with over a trillion parameters by using memory sharding in combination with data parallelism. DeepSpeed also provides a Pipeline Parallelism engine (inherited from the PipeDream research, see below) that can combine with ZeRO data parallelism, as well as many convenience features (gradient accumulation, automatic mixed precision, etc.). Importantly, DeepSpeed made these techniques accessible through relatively simple APIs and configuration files, so users could enable hybrid parallelism without extensive code changes. Galvatron's goals of easy integration and automation align with DeepSpeed's philosophy, and indeed Galvatron uses DeepSpeed under-the-hood for certain functions (memory optimization, initializing distributed training). The distinction is that DeepSpeed's hybrid parallelism is still largely static – e.g., one can run ZeRO stage 3 + pipeline parallel, but the degrees of each are set a priori. Galvatron could utilize DeepSpeed to set up a pipeline, then later decide to change the pipeline stages or switch parallelism, which is outside of what DeepSpeed alone offers. Thus, Galvatron can be seen as a higher-level orchestration that could even work on top of DeepSpeed, adding an AI-driven decision layer.

**5.3 Google GShard:** Lepikhin et al. (2020) developed GShard [5], a framework for scaling giant models (especially focusing on **Mixture-of-Experts (MoE) Transformers**) using **automatic sharding and conditional computation** [7]. GShard introduced a set of lightweight annotation APIs for model code that allow the XLA compiler to automatically shard computation and data across multiple devices. Using GShard, the authors trained a 600 billion-parameter multilingual translation model on 2048 TPU cores [15], demonstrating the power of combining data parallelism, model sharding, and expert parallelism (where only a subset of model parameters are active per input due to MoE). One of GShard's key benefits is it required minimal changes to the model code – the compiler took on the heavy lifting of parallelizing the model [30]. This is conceptually similar to Galvatron's ease-of-use goal. However, GShard's automatic sharding is largely compile-time (deciding a partitioning before training) and focused on a specific use case (very large MoE on TPUs). Galvatron, operating in Python on GPUs at runtime, tackles a broader scenario of dynamic adjustments and is not limited to XLA. Nonetheless, both aim to automate the distribution of computation with minimal manual intervention, reinforcing the trend toward compiler or runtime-based parallelism decisions rather than human decisions.

**5.4 Meta AI FairScale:** FairScale is a PyTorch extension library by Facebook (Meta) [8] that provides modular components for large-scale training, including Sharded Data Parallel (SDP) and Fully Sharded Data Parallel (FSDP). These techniques are similar to ZeRO: they shard model weights and gradients across processes to reduce memory usage, allowing bigger models or bigger batches on the same hardware. FairScale also includes a Pipeline parallel implementation and was used internally for training models like Blender (a large chatbot). FairScale's approach is more library-style; users manually mix and match components (for example, wrapping a model with fairscale.nn.FullyShardedDataParallel to shard it). FairScale makes hybrid parallel training more accessible in PyTorch, but again, the user must decide when to use FSDP vs data parallel, etc. Galvatron could incorporate ideas from FairScale, and indeed if Galvatron were to support PyTorch's native FSDP, it could programmatically toggle sharding strategies. The design philosophy of providing building blocks for efficient training is common to FairScale and Galvatron, but Galvatron's novelty is the dynamic coordination of those blocks by an intelligent manager. In effect, Galvatron could be seen as doing at runtime what an engineer might do manually with FairScale: experiment with sharding or data parallel degrees to see what runs fastest.

**5.5 Pipeline Parallelism Research (GPipe, PipeDream):** Two notable research works in pipeline parallelism are GPipe (Huang et al. 2019) and PipeDream (Narayanan et al. 2019) [27, 21]. GPipe introduced the idea of splitting a neural network into sequential partitions on different accelerators and using micro-batch pipelining with gradient accumulation to keep all partitions busy. GPipe achieved near-linear speedups for training very deep models by pipelining, albeit with the drawback of pipeline flush at epoch boundaries (since it used synchronous updates). PipeDream improved on this by allowing asynchronous pipeline parallelism where different pipeline stages could be training on different versions of the model (trades off some staleness for throughput). PipeDream also introduced mechanisms to allocate layers to stages in a balanced way and handle replication of stages to avoid bubbles. Both of these systems require deciding the pipeline partitioning before training (GPipe even needs it for the compiler). Galvatron's dynamic approach could, in theory, integrate pipeline partitioning algorithms from these works but apply them dynamically if a stage becomes a bottleneck. In fact, the example given in the documentation of Galvatron – reducing pipeline stages from 4 to 3 mid-training due to imbalance– echoes the goal of PipeDream to avoid idle stages. The difference is Galvatron does it adaptively rather than designing the pipeline schedule entirely upfront. Thus, Galvatron extends pipeline parallel research by making stage allocation and number a tunable parameter during the run.

**5.6 Other Hybrid Systems:** There are other efforts like **Mesh-TensorFlow** (Google) [24] which allowed model definitions to specify how tensors are split across a "mesh" of devices, enabling flexible parallelism for TPUs. And more recent developments such as **Alpa** (from Alibaba and others) [15] use cluster-of-gpu partitioning and even automated parallelism planning with cost models (Alpa can automatically choose a parallelization strategy for a given model graph using dynamic programming, somewhat similar in spirit to Galvatron's search but at compile time). These indicate a general direction in research: automatically find the best parallelization for a given model and hardware. Galvatron is unique in pushing this automation into the **runtime** of training, which allows it to adjust to observed performance, not just theoretical cost models.

# 6  Conclusion

We have presented a detailed analysis of **Galvatron**, a dynamic mixture-based parallel training framework that represents a significant step forward in distributed deep learning implementation. Galvatron (**Optimus-Megatron** in implementation) tackles one of the pain points of large-scale AI model training: the need to manually configure and fine-tune parallelism strategies for each model and hardware setup. By leveraging **dynamic parallelism**, Galvatron can **automate the selection**, **combination, and evolution of parallel training strategies** over the course of a single training run. This results in a system that is both highly efficient – extracting maximal performance from available GPUs – and user-friendly, lowering the barrier to entry for training massive models.

The technical contributions of Galvatron span the integration of profiling components, a strategy search algorithm, and a runtime reconfiguration mechanism built on top of robust libraries like Megatron-LM and DeepSpeed. In doing so, Galvatron doesn't reinvent the wheel for low-level operations, but rather innovates in the *control plane* of distributed training. This separation of concerns is elegant: it means future improvements in underlying frameworks (e.g., faster communication kernels or better optimizers) can be immediately leveraged by Galvatron, and likewise, Galvatron's ideas could potentially be integrated into those frameworks to enhance them with dynamic capabilities.

For practitioners, Galvatron offers a glimpse into the **future of large-scale training** – one where systems intelligently manage resources, and data scientists can focus on model design and data, not on parallelization details. As model sizes continue to grow and hardware becomes more heterogeneous, such dynamic systems will be crucial. We expect future work to extend Galvatron in various ways: more granular adaptation (maybe adjusting parallelism per layer group or per training epoch based on a learning schedule), incorporating new forms of parallelism (like expert parallelism for MoE models), and even handling dynamic resource pools (fault tolerance or elasticity). The foundation Galvatron provides is strong, and its open-source availability encourages the community to experiment and build upon it.

In conclusion, Galvatron demonstrates that combining multiple parallelism techniques with an automated, adaptive approach can yield superior results in training large neural networks [21]. It validates the hypothesis that no single static strategy is best for all situations – instead, a mixture of strategies orchestrated by intelligent software can consistently outpace human-configured setups. This work stands at the intersection of machine learning and systems research, showing how ideas from both domains (reinforcement learning, search algorithms, distributed systems, compilers) can come together to address one of the grand challenges of AI: training the next generation of models efficiently. Galvatron's implementation provides a practical tool for the community and a baseline for future innovations in this space.